\documentclass[
aps,prd,
12pt,
nopreprintnumbers,
showpacs,
eqsecnum,
nofootinbib
]{revtex4-1}

\usepackage{graphicx}
\usepackage{amssymb}

\def\Tr{{\rm Tr\,}}

\begin{document}

\title{Dynamical mass generation of spin-2 f{}ields in de Sitter space for
an $O(N)$ symmetric model at large $N$}
\author{Nahomi Kan}\email[]{kan@gifu-nct.ac.jp}
\affiliation{National Institute of Technology, Gifu College,
Motosu-shi, Gifu 501-0495, Japan}
\author{Kiyoshi Shiraishi}\email[]{shiraish@yamaguchi-u.ac.jp}
\affiliation{
Graduate School of Sciences and Technology for Innovation, Yamaguchi
University, Yamaguchi-shi, Yamaguchi 753--8512, Japan}
\date{\today}
\begin{abstract}
We consider the strong-coupling phase in a model of $O(N)$ spin-2 field
theory in de Sitter spacetime and the effective mass of spin-2 fields therein. In
the strong-coupling phase, the Higuchi bound
limits the mass parameter in the theory. The analysis using the large
$N$ approximation finds the critical value of the mass parameter with numerical
calculation.
\end{abstract}



\maketitle

\section{Introduction}
\label{introduction}

The massive spin-2 field theory has been studied for quite some time \cite{FP}.
However, in recent years, the theory of massive graviton
\cite{Hinterbichler,deRham} as a spin-2 particle has attracted attention as a
kind of modified gravitation theories. As a key to solving the cosmological
constant problem, it is also considered significant to investigate the
relationship between the mass of the graviton and a small cosmological constant.
Besides, bigravity theories \cite{SMvS} and multigravity theories are also
considered, and in special cases, it is also speculated that spin-2 particles
(not yet discovered) other than the graviton partly plays the role of dark
matter \cite{CCL}.

On the other hand, the existence of the Higuchi bound \cite{Higuchi1,Higuchi2} is
known for the mass of a spin-2 particle in maximally symmetric spacetime. It has
been shown that a negative norm state appears in the spin-2 field theory below the
critical mass-squared, $\frac{D-2}{D(D-1)}R$ (where $R$ is the
scalar curvature of the de Sitter space), and the theory becomes unstable (except
for the massless case). This is a distinctive feature of the spin-2 field theory
that is not seen in other spin fields. Therefore, in some sense, it is speculated
that the study of spin-2 theory is a very important key point related to both the
dark energy problem and the dark matter problem.

Now, for example, the masses of the (spin 1/2, 1) particles in the standard model
is obtained by the Higgs mechanism. On the other hand, how is the mass of spin-2
particles determined? In this paper, as a very bold assumption, we consider a
model in which the dynamical mass is determined from the interaction of $N$ spin-2
fields $h^a_{\mu\nu}$ $(a=1,\dots,N)$. 
Consider $N$ spin-2 fields $h_{\mu\nu}^a$
with $O(N)$ invariant interaction:
\begin{equation}
\mathcal{L}_{int}=-\frac{\lambda}{8N}(h^{a\,\mu\nu}h^a_{\mu\nu}-h^{a\mu}_\mu
h^{a\nu}_\nu)^2\,,
\end{equation}
where $\lambda$ is the coupling constant.
Then, the vacuum expectation value 
\begin{equation}
\langle\sigma
\rangle\equiv\langle m^2+
\frac{\lambda}{2N}(h^{a\,\mu\nu}h^a_{\mu\nu}-h^{a\mu}_\mu
h^{a\nu}_\nu)\rangle\,,
\end{equation}
(where $m$ is the mass parameter) is the mass squared of the spin-2 fields in the
strong-coupling vacuum.  For the spin-zero field model, the large $N$
approximation as a self-consistent approach has been devised since half a century
ago
\cite{CJP,KK,AKS,SAT}, which makes it possible to study the strong-coupling phase
relatively easily, and we also uses this method in this paper.%
\footnote{The large $N$ scalar field theory in de Sitter space has been studied in
Refs.~\cite{Serreau,GS,LR}, and recently, that in three
dimensional AdS space has been investigated in Ref.~\cite{BH}.} Since the mass in
the Lagrangian and the effective mass in the strong-coupling vacuum are
generally different, we are most interested in how the parameter restriction by
the Higuchi bound in de Sitter spacetime works.

Various no-go theories have been found for massless spin-2 field interactions
\cite{BDGH}. However, earlier discussions have already reported that the theory of
spin-2 with perfectly sound interactions requires an infinite number of
massive fields anyway
\cite{DPS,BHR}. The relationship between the string theory, which involves
infinite particle states, and the massive spin-2 theory has been investigated
\cite{LMMS}. 
Therefore, although the  model treated in the present paper does not have a number
of theoretic properties such as UV completeness, we have begun our study with the
expectation that the theory will have better properties in the emerging 
vacuum in nonperturvative regime.

In our model, defined in the next section, we proceed with the study
assuming that the $N$ spin-2 fields are independent of the graviton. However,
the mixing with the graviton is conceivable as a field of the same spin-2, so in
the future,  we will manage to construct an
effective model related to cosmology and elementary particle theory, after taking
the mixing properly. 
Of course, the existence of large-$N$ spin-2 fields with a degenerate mass 
is problematic in the literal sense. We expect that even in this simple model we
can see some features of the model with a finite number of fields.
We would like to leave such research as a subject
for future papers.

The structure of the present paper is as follows. In Sec.~\ref{sec2}, we introduce
the spin-2 model considered in this paper. In Sec.~\ref{sec3}, we consider the
case of four dimensional de Sitter spacetime. The critical value of the
renormalized Lagrangian mass, avoiding the Higuchi bound, is examined.
Finally, we discuss the conclusion and future prospects in the last section.
We have Appendix~\ref{AA} for the case of
three dimensional de Sitter spacetime.

\section{$O(N)$ symmetric spin-2 model in de Sitter spacetime}
\label{sec2}
Let us consider the following model.
The action of $N$ massive symmetric tensor field
$h^a_{\mu\nu}$ on a $D$-dimensional maximally symmetric spacetime with the constant
scalar curvature $R\equiv g^{\mu\nu}R_{\mu\nu}$, where the Ricci tensor
$R_{\mu\nu}$ equals to $\frac{R}{D}g_{\mu\nu}$, is written by
\cite{AD1,AD2,BKP,BGKP,BGP}
\begin{eqnarray}
S_{FP}&=&\int d^Dx\sqrt{-g}\Bigl[-\frac{1}{2}\nabla_\mu h^a_{\nu\rho}\nabla^\mu
h^{a\,\nu\rho}+
\nabla_\nu h^a_{\mu\rho}\nabla^\mu h^{a\,\nu\rho}-
\nabla_\mu h^a\nabla_\rho h^{a\,\mu\rho}+
\frac{1}{2}\nabla_\mu h^a\nabla^\mu h^a\nonumber \\
& &+\frac{R}{D}\Bigl(h^{a\,\mu\nu}h^a_{\mu\nu}-\frac{1}{2}h^ah^a\Bigr)
-\frac{1}{2}m_0^2(h^{a\,\mu\nu}h^a_{\mu\nu}-h^ah^a)\Bigr]\,,
\label{FP}
\end{eqnarray}
where $h\equiv h^\rho_\rho=g^{\mu\nu}h_{\mu\nu}$, $\nabla_\mu$ represents the
covariant derivative in terms of the metric $g_{\mu\nu}$, and
$m_0$ is the mass parameter (the bare Lagrangian mass). 
In addition, we assume the following $O(N)$ invariant interaction:
\begin{equation}
S_{int}=\int d^Dx\sqrt{-g}\,
\Bigl[-\frac{\lambda_0}{8N}(h^{a\,\mu\nu}h^a_{\mu\nu}-h^ah^a)^2\Bigr]\,,
\end{equation}
where $\lambda_0$ is the (bare) quartic self-interaction coupling constant.
In this paper,
we will not consider the possibility of $O(N)$ symmetry breaking, although it might
be interesting.

To introduce the auxiliary field $\sigma$, we add the following action, which is
nondynamical itself, for $\sigma$:
\begin{equation}
S_{aux}=\int d^Dx\sqrt{-g}\,
\left\{\frac{N}{2\lambda_0}\left[\sigma-m_0^2-\frac{\lambda_0}{2N}(h^{a\,\mu\nu}
h^a_{\mu\nu}-h^ah^a)\right]^2\right\}\,.
\end{equation}
Note that the functional integral over $\sigma$ is a trivial Gaussian integral.

It can be seen that the vacuum expectation value of $\sigma$ becomes the square of
the mass of the spin-2 fields. Performing a Gaussian integral over the spin-2
fields $h^a_{\mu\nu}$ (and the corresponding ghost fields with the gauge fixing) 
can then be performed in the partition function with the total action
$S=S_{FP}+S_{int}+S_{aux}$ and we obtain the one-loop effective action
up to an additive constant:
\begin{equation}
S_{eff}=\int d^Dx\sqrt{-g}\,\left[
\frac{N}{2\lambda_0}\sigma^2-\frac{Nm_0^2}{\lambda_0}\sigma-NL_0(\sigma)\right]
\equiv\int d^Dx\sqrt{-g}\,\Bigl[-NV(\sigma)\Bigr]\,,
\end{equation}
where \cite{DDLS,DLS,CD}
\begin{equation}
L_0(\sigma)\equiv\mathcal{V}_D^{-1}\left\{\frac{1}{2}\Tr\ln
\left[\Delta(1,1)-2\frac{R}{D}+\sigma\right]
-\frac{1}{2}\Tr\ln\left[\Delta(\mbox{$
\frac{1}{2},\frac{1}{2}$})-2\frac{R}{D}+\sigma\right]\right\}\,,
\end{equation}
with $\mathcal{V}_D$ is the volume of the spacetime, $\int\sqrt{-g}\,d^Dx$.
Here, the differential operators $\Delta(\mbox{$
\frac{1}{2},\frac{1}{2}$})$ and $\Delta(1,1)$ are defined as
\begin{equation}
\Delta(\mbox{$
\frac{1}{2},\frac{1}{2}$})\xi_\mu\equiv-\Box\xi_\mu+R_{\mu\nu}\xi^\nu\,,\quad
\Delta(1,1)\phi_{\mu\nu}\equiv-\Box\phi_{\mu\nu}+
R_{\mu\tau}\phi^\tau_{\nu}+
R_{\nu\tau}\phi^\tau_{\mu}-2R_{\mu\rho\nu\tau}\phi^{\rho\tau}\,,
\end{equation}
respectively, where $R_{\mu\rho\nu\tau}$ denotes the Riemann tensor.

The equation of motion $\frac{\delta S}{\delta\sigma}=0$, which determines the
extremum of the effective potential,
$\frac{dV}{d\sigma}=0$,
leads to
\begin{equation}
\frac{1}{\lambda_0}\sigma-\frac{m_0^2}{\lambda_0}-\frac{d
L_0(\sigma)}{d\sigma}=0\,,
\end{equation}
which gives the relation of the leading order in the expansion in $1/N$, thus it
is independent of $N$. The expected value of $\sigma$ can be
obtained by solving this equation.

We will compute the one-loop contribution of the spin-2 fields by with the spectrum
of the Laplacian on a $D$-sphere $S^D$, as the Euclidean version of
$D$-dimensional de Sitter spacetime. Hereafter, we set the constant curvature of
the space as
\begin{equation}
R_{\mu\rho\nu\tau}=\frac{\Lambda}{D-1}(g_{\mu\nu}g_{\rho\tau}-
g_{\mu\tau}g_{\rho\nu})\,,\quad R_{\mu\nu}=g^{\rho\tau}R_{\mu\rho\nu\tau}=\Lambda
g_{\mu\nu}\,,\quad\mbox{and}
\quad R=D\Lambda\,,
\end{equation}
where $\Lambda$ is a positive cosmological constant, while the volume is given by
\begin{equation}
\mathcal{V}_D=\frac{2\pi^{\frac{D+1}{2}}}{\Gamma\left(\frac{D+1}{2}\right)}
\left(\frac{\Lambda}{D-1}\right)^{-D/2}\,.
\end{equation}


Then, the first derivative of $L_0$ is formally given by
\cite{DDLS,DLS,CD,RO1,RO2,Higuchi,Allen,MM}
\begin{eqnarray}
\frac{dL_0(\sigma)}{d\sigma}&=&
\frac{\Gamma\left(\frac{D+1}{2}\right)}{4\pi^{\frac{D+1}{2}}}
\left(\frac{\Lambda}{D-1}\right)^{D/2}\times \nonumber \\
& &
\left\{\sum_{\ell=0}^\infty d_2(\ell)\left[
\frac{\Lambda}{D-1}(\ell+2)(\ell+D+1)+\sigma
\right]^{-1}\right.\nonumber \\
& &\left.
-\sum_{\ell=0}^\infty d_1(\ell)\left[
\frac{\Lambda}{D-1}\ell(\ell+D+1)+\sigma\right]^{-1}\right\}\,,
\label{dL}
\end{eqnarray}
where the degeneracies are
\begin{equation}
d_2(\ell)\equiv\frac{(D+1)(D-2)}{2}(\ell+1)(\ell+D+2)(2\ell+D+3)
\frac{(\ell+D-1)!}{(D-1)!(\ell+3)!}\,,
\end{equation}
and
\begin{equation}
d_1(\ell)\equiv(\ell+1)(\ell+D)(2\ell+D+1)
\frac{(\ell+D-2)!}{(D-2)!(\ell+2)!}\,.
\end{equation}

Note that $\frac{dL_0}{d\sigma}$ contains
divergences, which must be addressed by parameter renormalization. 
We will show concrete methods and numerical results for four dimensional
spacetime in the next
section,  and the results for three dimensional spacetime 
in Appendix~\ref{AA}.



\section{four dimensions ($D=4$)}
\label{sec3}

\subsection{Calculation of one-loop contribution}

For four dimensions, the expression (\ref{dL}) becomes%
\footnote{In the limit of $\Lambda\rightarrow 0$, we find that
$\frac{dL_0(\sigma)}{d\sigma}=\frac{1}{8\pi^2}\int\frac{k^3dk}{k^2+\sigma}
=\frac{g}{2}\int\frac{d^4k}{(2\pi)^4}\frac{1}{k^2+\sigma}$ with $g=2$. According to
\cite{DDLS}, the absence of the discontinuity between the massive case and
massless case at the one-loop level has been noted
(thus, $g=2$ instead of $g=5$).}
\begin{equation}
\frac{dL_0(\sigma)}{d\sigma}=
\frac{\Lambda}{16\pi^{2}}
\left\{\sum_{\ell=0}^\infty d_2(\ell)\left[
(\ell+2)(\ell+5)+\frac{3\sigma}{\Lambda}
\right]^{-1}
-\sum_{\ell=0}^\infty d_1(\ell)\left[
\ell(\ell+5)+\frac{3\sigma}{\Lambda}\right]^{-1}\right\}\,,
\end{equation}
where
\begin{equation}
d_2(\ell)=
\frac{5}{6}(\ell+1)(\ell+6)(2\ell+7)\quad
\mbox{and}\quad
d_1(\ell)=
\frac{1}{2}(\ell+1)(\ell+4)(2\ell+5)\,.
\end{equation}

Using the integration formula 
\begin{equation}
\frac{1}{z^2-\beta^2}=\int_0^\infty e^{-zt}\frac{\sinh \beta t}{\beta}dt\,,
\end{equation}
$\frac{dL_0}{d\sigma}$ is expressed as
\begin{eqnarray}
\frac{dL_0(\sigma)}{d\sigma}&=&
\frac{\Lambda}{16\pi^{2}}
\left\{\frac{5}{16}\int_\epsilon^\infty 
\frac{e^{-\frac{7}{2}t}(2-7e^{t}+7e^{2t})}{\left(\sinh\frac{t}{2}\right)^4}
\frac{\sinh \beta_2(\sigma) t}{\beta_2(\sigma)}dt
\right.\nonumber \\
& &\left.
-\frac{1}{16}\int_\epsilon^\infty 
\frac{e^{-\frac{5}{2}t}(1-5e^{t}+10e^{2t})}{\left(\sinh\frac{t}{2}\right)^4}
\frac{\sinh \beta_1(\sigma) t}{\beta_1(\sigma)}dt\right\}\,,
\end{eqnarray}
where
\begin{equation}
\beta_2(\sigma)\equiv\sqrt{\frac{9}{4}-\frac{3\sigma}{\Lambda}}\quad
\mbox{and}\quad
\beta_1(\sigma)\equiv\sqrt{\frac{25}{4}-\frac{3\sigma}{\Lambda}}\,.
\end{equation}
Here, we introduced a UV regulator $\epsilon$ in order to avoid divergences at the
lower limit in the integrals.

Now, noting that
\begin{equation}
\frac{\sinh\beta t}{\beta}=t+\frac{\beta^2}{6}t^3+O(t^5)\,,
\end{equation}
we renormalize the divergent integral by defining the renormalized
parameters as follows:%
\footnote{We adopt a kind of ``minimal subtraction'' of divergences in the integral
in this paper.}
\begin{eqnarray}
\frac{m^2}{\lambda}&=&\frac{m_0^2}{\lambda_0}+
\frac{\Lambda}{16\pi^{2}}
\left\{\frac{5}{16}\int_\epsilon^\infty 
\frac{e^{-\frac{7}{2}t}(2-7e^{t}+7e^{2t})}{\left(\sinh\frac{t}{2}\right)^4}
\left(t+\frac{3}{8}t^3\right)dt\right.\nonumber \\
& &\left.
-\frac{1}{16}\int_\epsilon^\infty 
\frac{e^{-\frac{5}{2}t}(1-5e^{t}+10e^{2t})}{\left(\sinh\frac{t}{2}\right)^4}
\left(t+\frac{25}{24}t^3\right)dt\right\}\,,
\end{eqnarray}
and
\begin{eqnarray}
\frac{1}{\lambda}&=&\frac{1}{\lambda_0}+
\frac{1}{32\pi^{2}}
\left\{\frac{5}{16}\int_\epsilon^\infty 
\frac{e^{-\frac{7}{2}t}(2-7e^{t}+7e^{2t})}{\left(\sinh\frac{t}{2}\right)^4}
t^3dt\right.\nonumber
\\ & &\left.
-\frac{1}{16}\int_\epsilon^\infty 
\frac{e^{-\frac{5}{2}t}(1-5e^{t}+10e^{2t})}{\left(\sinh\frac{t}{2}\right)^4}
t^3dt\right\}\,.
\end{eqnarray}
Then, the equation of motion for $\sigma$ (the gap equation) is rewritten as
\begin{equation}
\frac{1}{\lambda}\sigma-\frac{m^2}{\lambda}-\frac{dL(\sigma)}{d\sigma}=0\,,
\end{equation}
where
\begin{equation}
\frac{dL(\sigma)}{d\sigma}\equiv
\frac{dL_0(\sigma)}{d\sigma}-\left(\frac{m^2}{\lambda}-
\frac{m_0^2}{\lambda_0}\right)+\left(\frac{1}{\lambda}-
\frac{1}{\lambda_0}\right)\sigma\,.
\end{equation}
After the renormalization, $\frac{dL(\sigma)}{d\sigma}$ is calculable
by the integral by setting $\epsilon\rightarrow 0$.


\begin{figure}[ht]
\centering
\includegraphics
{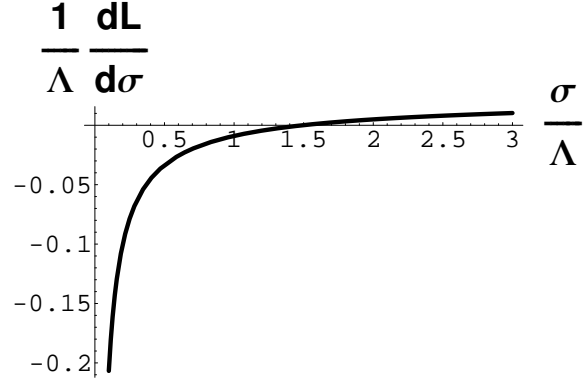}
\caption{$\frac{1}{\Lambda}\frac{dL}{d\sigma}$ for the four dimensional case as the
function of
$\frac{\sigma}{\Lambda}$.}
\label{fig1}
\end{figure}

We show $\frac{1}{\Lambda}\frac{dL}{d\sigma}$ as the function of
$\frac{\sigma}{\Lambda}$ in Fig.~\ref{fig1}, which is obtained by numerical
integration in the case of four dimensions. 
We should notice that $\frac{dL}{d\sigma}$ is calculable without regard to the
Higuchi bound for
$\sigma>0$. It can be seen that 
$L$ is convex downward as a function of $\sigma$,
since $\frac{1}{\Lambda}\frac{dL}{d\sigma}$ is monotonously increasing
as shown in Fig.~\ref{fig1}.
The minimum of
$L$ is located at $\sigma/\Lambda\approx 1.5$.

\subsection{Negative coupling constant $\lambda$}

For the $O(N)$ scalar theory, it is reported that there appears the $O(N)$
symmetric stable ground state when the renormalized coupling $\lambda$ is negative
\cite{KK,AKS,SAT}. Therefore, we first examine the case with $\lambda<0$
in our model.%
\footnote{Most recently, a preprint \cite{Weller} has appeared. The preprint
includes a recent discussion on the negative coupling constant in scalar field
theory and some important references.}

\begin{figure}[ht]
\centering
\includegraphics
{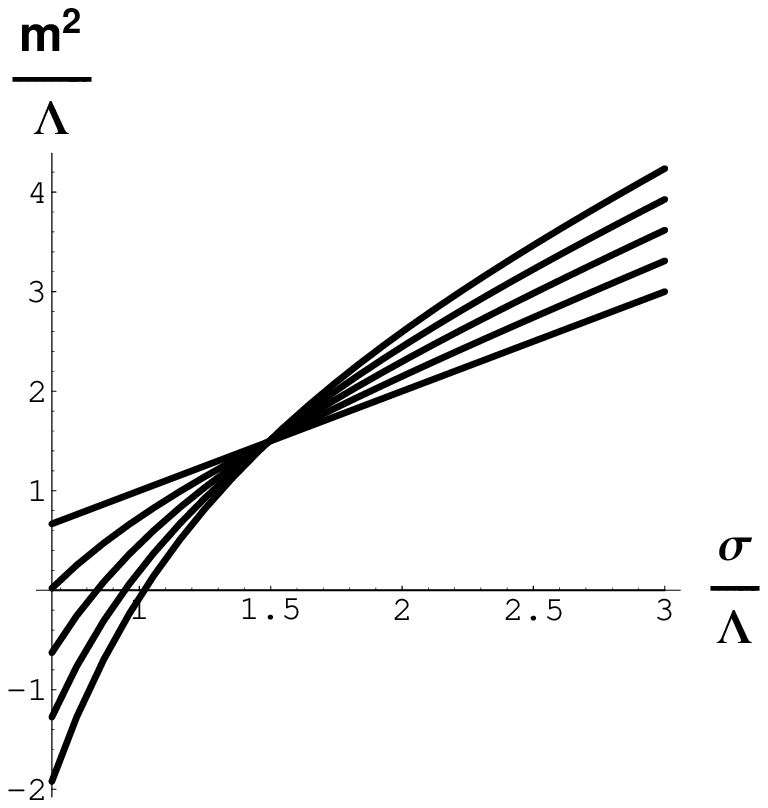}
\caption{The relation between $\frac{m^2}{\Lambda}$
and $\frac{\sigma}{\Lambda}$ for some specific values of the coupling
$\lambda(<0)$. The curves correspond to $|\lambda|=0,30,60,90,120$, 
whose leftmost points line up from top to bottom.}
\label{fig2}
\end{figure}

\begin{figure}[ht]
\centering
\includegraphics
{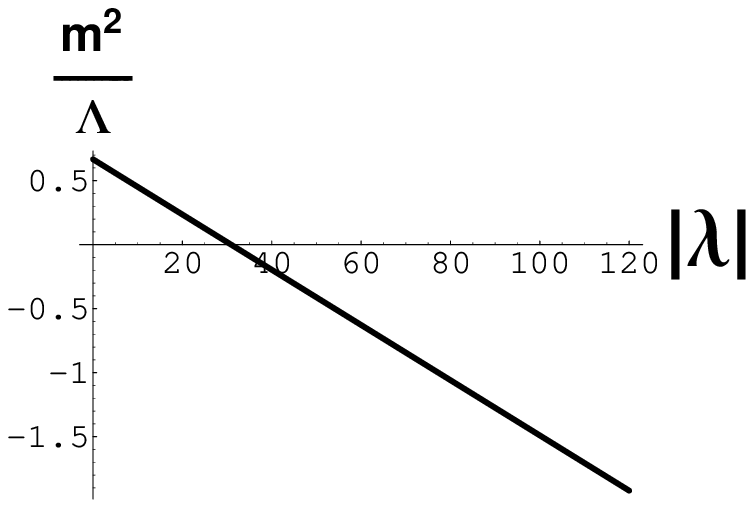}
\caption{The lower limit of $m^2/\Lambda$ for $\sigma/\Lambda>2/3$,
as a function of $|\lambda|$.}
\label{fig3}
\end{figure}

Since the equation of motion can
be read as, for $\lambda<0$,
\begin{equation}
\frac{m^2}{\Lambda}=\frac{\sigma}{\Lambda}+|\lambda|\frac{1}{\Lambda}\frac{d
L}{d\sigma}\qquad(\lambda<0)\,,
\label{310}
\end{equation}
we can exhibit the relation between $\frac{m^2}{\Lambda}$
and the vacuum expectation value of $\frac{\sigma}{\Lambda}$ for some specific
values of the coupling
$\lambda$
 in Fig.~\ref{fig2} (where we use the simple $\sigma$ instead of
$\langle\sigma\rangle$ for the vacuum expectation value). We show the
region where
$\sigma/\Lambda>2/3$, above the Higuchi bound \cite{Higuchi1,Higuchi2}. 
The lower limit of $m^2/\Lambda$ as a function of $|\lambda|$, which yields
$\sigma/\Lambda>2/3$, is shown in Fig.~\ref{fig3}.

\subsection{Positive coupling constant $\lambda$}

For $\lambda>0$, the equation of motion for $\sigma$ reads
\begin{equation}
\frac{m^2}{\Lambda}=\frac{\sigma}{\Lambda}-\lambda\frac{1}{\Lambda}\frac{d
L}{d\sigma}\qquad(\lambda>0)\,.
\end{equation}

\begin{figure}[ht]
\centering
\includegraphics
{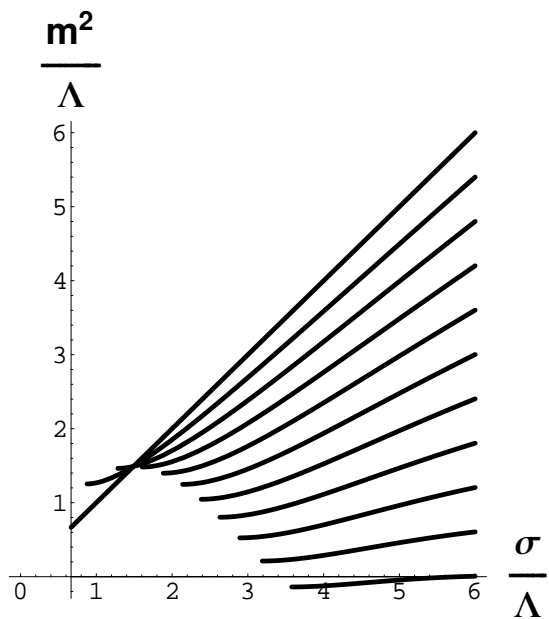}
\caption{The relation between $\frac{m^2}{\Lambda}$
and $\frac{\sigma}{\Lambda}$ for some specific values of the coupling
$\lambda(>0)$. The curves correspond to
$\lambda=0,30,60,90,120,150,180,210,240,270,300$,  whose leftmost points line up
from left to right.}
\label{fig4}
\end{figure}

\begin{figure}[ht]
\centering
\includegraphics
{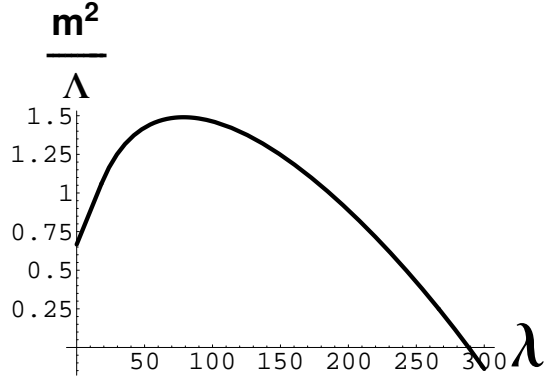}
\caption{The lower limit of $m^2/\Lambda$ for $\sigma/\Lambda>2/3$,
as a function of $\lambda$.}
\label{fig5}
\end{figure}

We show the relation between $\frac{m^2}{\Lambda}$
and the vacuum expectation value of $\frac{\sigma}{\Lambda}$ for some specific
values of the coupling $\lambda$ in Fig.~\ref{fig4}. We find the
region where
$\sigma/\Lambda>2/3$, above the Higuchi bound \cite{Higuchi1,Higuchi2}.
The increasing lines in the figure correspond to the extremum of the action,
which is continuously connected to that in the limit of $\lambda\rightarrow 0$. 
The lower limit of
$m^2/\Lambda$ as a function of
$\lambda$, which yields
$\sigma/\Lambda>2/3$, is shown in Fig.~\ref{fig5}.

In the present analysis, we assume that $O(N)$ symmetry is unbroken.
However, especially in the case of $m^2<0$, which realizes for $\lambda>\approx
289$, the unbroken $O(N)$ symmetry is problematic as in the case with the classical
Higgs potential.
More detailed analysis, such as Ref.~\cite{SAT} for scalar theory, is needed to
clarify the symmetry breaking in our present model.
However, since our model deals with fields with spacetime indices, similar
analyses of Ref.~\cite{SAT} must be reconstructed, and further research is left as
a future topic.

\section{Conclusion and outlook}
\label{conclusion}

In this paper, in the $O(N)$ spin-2 model, the effective mass in a strong-coupling
vacuum is studied by the large $N$ approximation, and the critical values of the
mass parameter, which exceeds the Higuchi bound, are numerically estimated. 
If we take a large enough $m^2$, we can
find a strong-coupling vacuum with an effective mass that exceeds the Higuchi
bound. 

Furthermore, for sufficiently large coupling constant, it is possible that the
effective mass is created when the mass parameter is zero (if $|\lambda|>31$
for $\lambda<0$ and if $\lambda>289$ for $\lambda>0$, in the case with $D=4$). 
There is also the well-known problem of discontinuity
\cite{Hinterbichler,deRham,DV} at
$m=0$,%
\footnote{Another discontinuity is also found at $m^2=\frac{D-2}{D-1}\Lambda$
\cite{Hinterbichler,deRham,DLS}.}
 so this point may require additional study in the future, though the
absence of the discontinuity at one-loop level is reported in Ref.~\cite{DDLS}.

The bound has been obtained in our toy
model, but we hope it makes some sense in the choice of models in realistic models
(including an infinite number of fields). Also, in such a case, the quantum effects
of other matter fields should inevitably come into play. After incorporating them,
we would like to consider the running of coupling constants and possible phase
transitions.
The stability of the vacuum is the most important topic that should be investigated
further. 

In this paper, we have also presented expressions in general $D$ dimensions, but 
it may be considered that renormalization or evaluation of parameters in higher
dimensions should be done more carefully than in four dimensions, so we
would like to leave it as a future task. However, like the calculation of quantum
effects in higher dimensional theory (e.g., \cite{CW}), it may be possible to
perform numerical calculations in a similar model in odd-dimensional spacetime. We
would like to consider such a issue in the future. 
We assumed that the
$O(N)$ symmetry is not broken in this paper (for simplicity as in the initial
stage of the study), but we would like to consider the case where $O(N)$ is broken
in the future. We wish to examine the model when the background spacetime is
also the Nariai spacetime \cite{Nariai}, a higher-dimensional Kaluza--Klein
spacetime, etc. Paying attention to whether the anisotropy can be avoided or
favored, we want to study their consequences. In the future, we would like to
research bold hypotheses such as the self-consistent de Sitter universe (e.g.,
Refs.~\cite{LR,WA,AW}) in the strong-coupling phase.




\appendix


\section{THREE DIMENSIONS ($D=3$)}
\label{AA}

\subsection{Calculation of one-loop contribution}
As in the four dimensional case, we find the following.
The formal expression of the one-loop calculation reads, for $D=3$,
\begin{equation}
\frac{dL_0}{d\sigma}=
\frac{\sqrt{\Lambda}}{4\sqrt{2}\pi^{2}}
\left\{\sum_{\ell=0}^\infty d_2(\ell)\left[
(\ell+2)(\ell+4)+\frac{2\sigma}{\Lambda}
\right]^{-1}
-\sum_{\ell=0}^\infty d_1(\ell)\left[
\ell(\ell+4)+\frac{2\sigma}{\Lambda}\right]^{-1}\right\}\,,
\end{equation}
where
\begin{equation}
d_2(\ell)=
2(\ell+1)(\ell+5)\quad
\mbox{and}\quad
d_1(\ell)=
2(\ell+1)(\ell+3)\,.
\end{equation}
As the treatment in the previous section, we can find its integration form:
\begin{eqnarray}
\frac{dL_0(\sigma)}{d\sigma}&=&
\frac{\sqrt{\Lambda}}{4\sqrt{2}\pi^{2}}
\left\{\frac{1}{4}\int_\epsilon^\infty 
\frac{e^{-\frac{5}{2}t}(-3+5e^{t})}{\left(\sinh\frac{t}{2}\right)^3}
\frac{\sinh \beta_2(\sigma) t}{\beta_2(\sigma)}dt\right.\nonumber \\
& &\left.
-\frac{1}{4}\int_\epsilon^\infty 
\frac{e^{-\frac{3}{2}t}(-1+3e^{t})}{\left(\sinh\frac{t}{2}\right)^3}
\frac{\sinh \beta_1(\sigma) t}{\beta_1(\sigma)}dt\right\}\,,
\end{eqnarray}
where
\begin{equation}
\beta_2(\sigma)\equiv\sqrt{1-\frac{2\sigma}{\Lambda}}\quad
\mbox{and}\quad
\beta_1(\sigma)\equiv\sqrt{4-\frac{2\sigma}{\Lambda}}\,.
\end{equation}

The renormalization should be done as
\begin{equation}
\frac{m^2}{\lambda}=\frac{m_0^2}{\lambda_0}+
\frac{\sqrt{\Lambda}}{16\sqrt{2}\pi^{2}}
\left\{\int_\epsilon^\infty 
\frac{e^{-\frac{5}{2}t}(-3+5e^{t})}{\left(\sinh\frac{t}{2}\right)^3}
t dt-\int_\epsilon^\infty 
\frac{e^{-\frac{3}{2}t}(-1+3e^{t})}{\left(\sinh\frac{t}{2}\right)^3}
tdt\right\}\,,
\end{equation}
and the coupling constant $\lambda_0=\lambda$ does not
undergo renormalization correction. Then, we set
\begin{equation}
\frac{dL(\sigma)}{d\sigma}\equiv
\frac{dL_0(\sigma)}{d\sigma}-\left(\frac{m^2}{\lambda}-
\frac{m_0^2}{\lambda_0}\right)\,.
\end{equation}

\begin{figure}[ht]
\centering
\includegraphics
{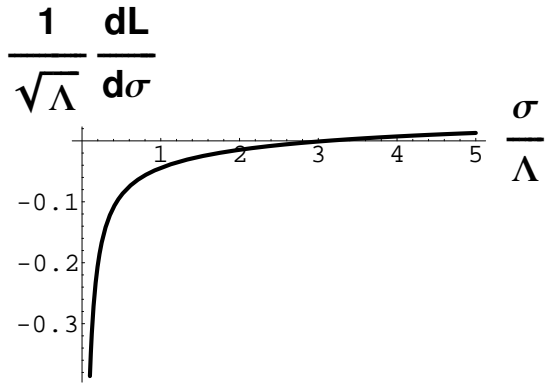}
\caption{$\frac{1}{\sqrt{\Lambda}}\frac{dL}{d\sigma}$ in the three dimensional
case as the function of
$\frac{\sigma}{\Lambda}$.}
\label{fig6}
\end{figure}

Qualitatively, it is the same as in the case of four dimensions. Note, however,
that the dimensions of the coupling constants are different.
We show $\frac{1}{\sqrt{\Lambda}}\frac{dL}{d\sigma}$ as the function of
$\frac{\sigma}{\Lambda}$ in Fig.~\ref{fig6} in the case of three dimensions. It can
be seen that
$L$ is convex downward as a function of $\sigma$ as in the four dimensional case.
The minimum of $L$ is located
at $\sigma/\sqrt{\Lambda}\approx 3$.

\subsection{Negative coupling constant $\lambda$}

\begin{figure}[ht]
\centering
\includegraphics
{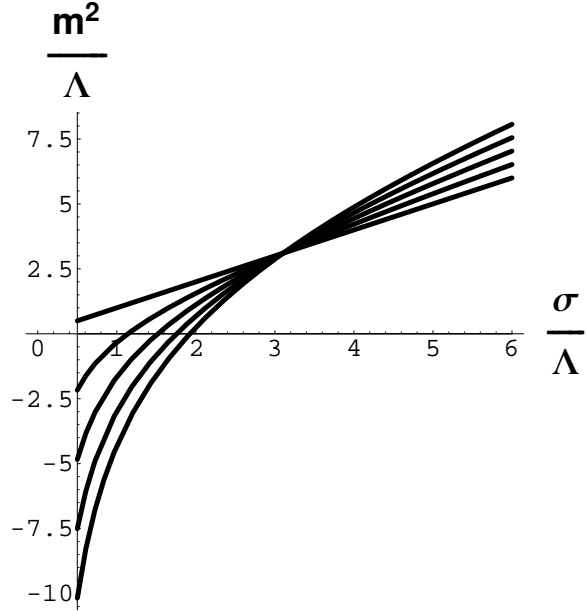}
\caption{the relation between $\frac{m^2}{\Lambda}$
and $\frac{\sigma}{\Lambda}$ for some specific values of the coupling
$\lambda(<0)$. The lines correspond to
$|\lambda|/\sqrt{\Lambda}=0,30,60,90,120$, whose leftmost points line up from top
to bottom.}
\label{fig7}
\end{figure}

The equation of motion can be read as, for $\lambda<0$,
\begin{equation}
\frac{m^2}{\Lambda}=\frac{\sigma}{\Lambda}+\frac{|\lambda|}{\sqrt{\Lambda}}\frac{1}{\sqrt{\Lambda}}
\frac{dL}{d\sigma}\,,
\end{equation}
where $\frac{|\lambda|}{\sqrt{\Lambda}}$ is dimensionless.
We show the relation between $\frac{m^2}{\Lambda}$
and $\frac{\sigma}{\Lambda}$ for some specific values of
$\frac{|\lambda|}{\sqrt{\Lambda}}$
 in Fig.~\ref{fig7}.  
We show the region where $\sigma/\Lambda>1/2$, which is above the Higuchi
bound in three dimensions. 

\begin{figure}[ht]
\centering
\includegraphics
{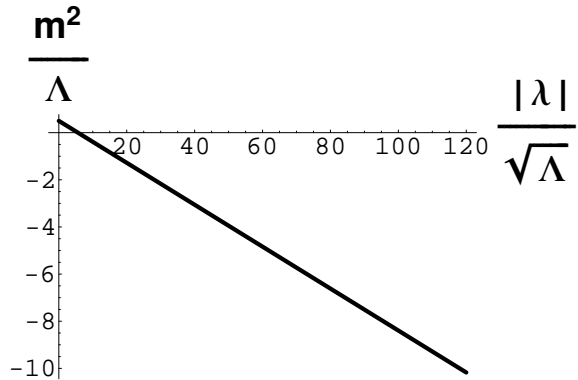}
\caption{The lower limit of $m^2/\Lambda$ for $\sigma/\Lambda>1/2$, as a
function of $|\lambda|/\sqrt{\Lambda}$.}
\label{fig8}
\end{figure}

The lower limit of the mass
parameter, which yields $\sigma/\Lambda>1/2$, is shown in Fig.~\ref{fig8}.

\subsection{Positive coupling constant $\lambda$}

\begin{figure}[ht]
\centering
\includegraphics
{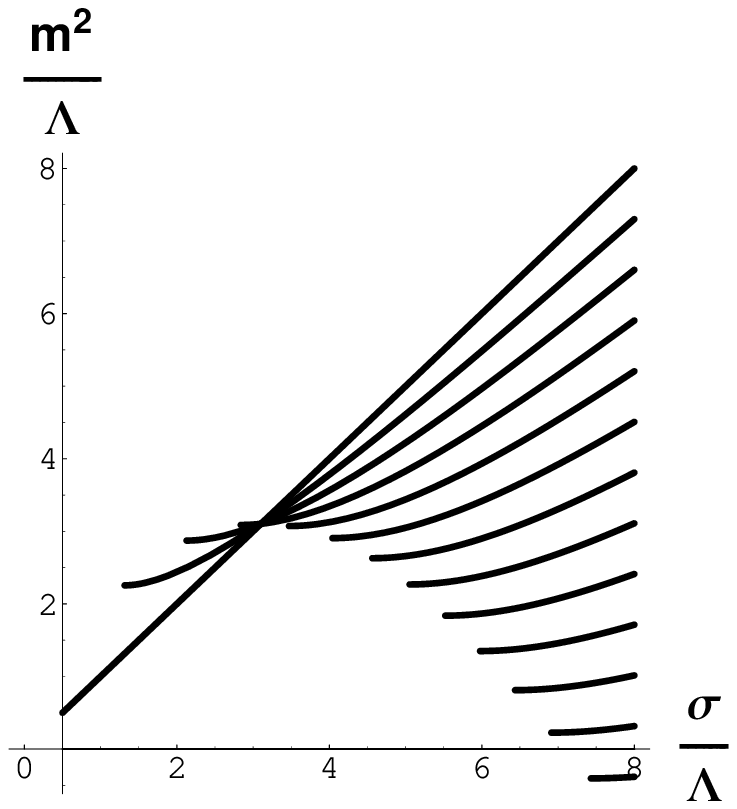}
\caption{the relation between $\frac{m^2}{\Lambda}$
and $\frac{\sigma}{\Lambda}$ for some specific values of the coupling
$\lambda(>0)$. The lines correspond to
$\lambda/\sqrt{\Lambda}=0,30,60,90,120,150,180,210,240,270,300,330,360$, whose
leftmost points line up from left to right.}
\label{fig9}
\end{figure}

The equation of motion can be read as, for $\lambda>0$,
\begin{equation}
\frac{m^2}{\Lambda}=\frac{\sigma}{\Lambda}-\frac{\lambda}{\sqrt{\Lambda}}\frac{1}{\sqrt{\Lambda}}
\frac{dL}{d\sigma}\,.
\end{equation}
We show the relation between $\frac{m^2}{\Lambda}$
and $\frac{\sigma}{\Lambda}$ for some specific values of
$\frac{\lambda}{\sqrt{\Lambda}}$
 in Fig.~\ref{fig9}.  
We show the region where $\sigma/\Lambda>1/2$, which is above the Higuchi
bound in three dimensions. 

\begin{figure}[ht]
\centering
\includegraphics
{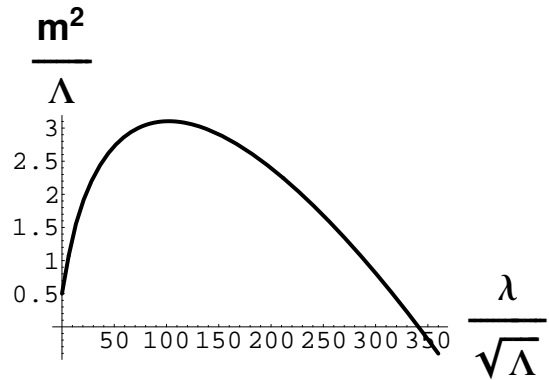}
\caption{The lower limit of $m^2/\Lambda$ for $\sigma/\Lambda>1/2$, as a
function of $\lambda/\sqrt{\Lambda}$.}
\label{fig10}
\end{figure}

The lower limit of the mass
parameter, which yields $\sigma/\Lambda>1/2$, is shown in Fig.~\ref{fig10}.

\acknowledgments
The authors thank the anonymous referees for pointing out an error in the early
manuscript and the importance of the negative self-coupling constant.



\bibliographystyle{apsrev4-1}

\begin{thebibliography}{99}
\bibitem{FP}
M.~Fierz and W.~Pauli,
``On relativistic wave function for particles of arbitrary spin in an
electromagnetic field'', Proc. R. Lond. \textbf{A173} (1939) 211. 
\bibitem{Hinterbichler}
K.~Hinterbichler,
``Theoretical aspects of massive gravity'',
Rev. Mod. Phys. \textbf{84} (2012) 671. 
\bibitem{deRham}
C.~de Rham,
``Massive gravity'',
Living Rev. Rel. \textbf{17} (2014) 7. 
\bibitem{SMvS}
A.~Schmidt-May and M.~von Strauss,
``Recent developments in bimetric theory'',
J. Phys. \textbf{A49} (2016) 183001.
\bibitem{CCL} H.~Cai, G.~Cacciapaglia and S.~J.~Lee,
``Massive gravitons as feebly interacting dark matter candidates'', 
Phys. Rev. Lett. \textbf{128} (2023) 081806 (6 pages).
\bibitem{Higuchi1}
A.~Higuchi,
``Forbidden mass range for spin-2 field theory in de Sitter space-time'',
Nucl. Phys. \textbf{B282} (1987) 397. 
\bibitem{Higuchi2}
A.~Higuchi,
``Massive symmetric tensor field theory in space-times with a positive
cosmological constant'', 
Nucl. Phys. \textbf{B325} (1989) 745. 
\bibitem{CJP} S.~Coleman, R.~Jackiw and H.~D.~Politzer,
``Spontaneous symmetry breaking in the $O(N)$ model for large $N$'',
Phys. Rev. \textbf{D10} (1974) 2491. 
\bibitem{KK} M.~Kobayashi and T.~Kugo,
``On the ground state of $O(N)$-$\lambda\phi^4$ model'',
Prog. Theor. Phys. \textbf{54} (1975) 1537. 
\bibitem{AKS} L.~F.~Abbott, J.~S.~King and H.~J.~Schnitzer,
``Bound states, tachyons, and restoration of symmetry in the $1/N$ expansion'',
Phys. Rev. \textbf{D13} (1976) 2212. 
\bibitem{SAT} P.~M.~Stevenson, B.~All\`es and R.~Tarrach,
``$O(N)$-symmetric $\lambda\phi^4$ theory: The Gaussian-effective-potential
approach'', 
Phys. Rev. \textbf{D35} (1987) 2407. 
\bibitem{Serreau} J.~Serreau,
``Effective potential for quantum fields in a de Sitter geometry'',
Phys. Rev. Lett. \textbf{107} (2011) 191103. 
\bibitem{GS} F.~Gautier and J.~Serreau,
``Scalar field correlator in de Sitter space at next-to-leading order in a $1/N$
expansion'', 
Phys. Rev. \textbf{D92} (2015) 105035. 
\bibitem{LR} D.~L.~L\'opez Nacir and J.~Rovner,
``Quantum backreaction of $O(N)$-symmetric scalar fields and de Sitter
spacetimes at the renormalization point: Renormalization schemes and
the screening of the cosmological constant'', 
Phys. Rev. \textbf{D103} (2021) 125002. 
\bibitem{BH} P.~Basu, S.~Haridev and P.~Samantray,
``On aspects of spontaneous symmetry breaking in Rindler and anti-de Sitter
spacetimes for the $O(N)$ linear sigma model'', 
Phys. Rev. \textbf{D107} (2023) 105004. 
\bibitem{BDGH} N.~Boulanger, T.~Damour, L.~Gualtieri and M.~Henneaux,
``Inconsistency of interacting, multi-graviton theories'', 
Nucl. Phys. \textbf{B597} (2001) 127. 
\bibitem{DPS}
M.~J.~Duff, C.~N.~Pope and K.~S.~Stelle,
``Consistent interacting massive spin-2 requires an infinity of states'',
Phys. Lett. \textbf{B223} (1989) 386. 
\bibitem{BHR} J.~Bonifacio, K.~Hinterbichler and R.~A.~Rosen,
``Constraints on a gravitational Higgs mechanism'', 
Phys. Rev. \textbf{D100} (2019) 084017 (13 pages).
\bibitem{LMMS} 
D.~L\"ust, C. Markou, P.~Mazloumi and S.~Stieberger,
``Extracting bigravity from string gravity'',
JHEP \textbf{2112} (2021) 220.
\bibitem{AD1}
C.~Aragone and S.~Deser,
``Constraints on gravitationally coupled tensor fields'',
Nuovo Cim. \textbf{A3} (1971) 709. 
\bibitem{AD2}
C.~Aragone and S.~Deser,
``Consistency problems of spin-2-gravity coupling'',
Nuovo Cim. \textbf{B57} (1980) 33. 
\bibitem{BKP}
I.~L.~Buchbinder, V.~A.~Krykhtin and V.~D.~Pershin,
``Consistent equations for massive spin-2 field coupled to gravity in string
theory'', 
Phys. Lett. \textbf{B466} (1999) 216. 
\bibitem{BGKP}
I.~L.~Buchbinder, D.~M.~Gitman, V.~A.~Krykhtin and V.~D.~Pershin,
``Equation of motion for massive spin-2 field coupled to gravity'',
Nucl. Phys. \textbf{B584} (2000) 615. 
\bibitem{BGP}
I.~L.~Buchbinder, D.~M.~Gitman and V.~D.~Pershin,
``Causality of massive spin-2 field in external gravity'',
Phys. Lett. \textbf{B492} (2000) 161. 
\bibitem{DDLS} F.~A.~Dilkes, M.~J.~Duff, J.~T.~Liu and H.~Sati,
``Quantum discontinuity between zero and infinitesimal graviton mass with a
$\Lambda$ term'', 
Phys. Rev. Lett. \textbf{87} (2001) 041301 (4 pages).
\bibitem{DLS} M.~J.~Duff, J.~T.~Liu and H.~Sati,
``Quantum $M^2\rightarrow 2\Lambda/3$ discontinuity for massive gravity with a
$\Lambda$ term'', 
Phys. Lett. \textbf{B516} (2001) 156. 
\bibitem{CD} S.~M.~Christensen and M.~J.~Duff,
``Quantizing gravity with a cosmological constant'',
Nucl. Phys. \textbf{B170} (1980) 480. 
\bibitem{RO1} M.~A.~Rubin and C.~R.~Ord\'o\~nez,
``Eigenvalues and degeneracies for $n$-dimensional tensor spherical harmonics'', 
J. Math. Phys. \textbf{25} (1984) 2888.
\bibitem{RO2} M.~A.~Rubin and C.~R.~Ord\'o\~nez,
``Symmetric-tensor eigenspectrum of the Laplacian on $n$-spheres'',  
J. Math. Phys. \textbf{26} (1985) 65.
\bibitem{Higuchi} A.~Higuchi,
``Symmetric tensor spherical harmonics on the $N$-sphere and their application
to the de Sitter group $SO(N,1)$'',  
J. Math. Phys. \textbf{28} (1987) 1553.
J. Math. Phys. \textbf{43} (2002) 6385(E).
\bibitem{Allen} B.~Allen,
``Phase transitions in de Sitter space'',
Nucl. Phys. \textbf{B226} (1983) 228. 
\bibitem{MM} P.~Mazur and E.~Mottola,
``Spontaneous breaking of de Sitter symmetry by radiative effects'',
Nucl. Phys. \textbf{B278} (1986) 694. 
\bibitem{Weller} R.~D.~Weller,
``Can negative bare couplings make sense? The $\vec{\phi}^4$ theory at large
$N$'',  arXiv:2310.02516 [hep-th].
\bibitem{DV} H.~van Dam and M.~Veltman,
``Massive and massless Yang--Mills and gravitational fields'', 
Nucl. Phys. \textbf{B22} (1970) 397. 
\bibitem{CW} P.~Candelas and S.~Weinberg,
``Calculations of gauge couplings and compact circumferences from
self-consistent dimensional reduction'', 
Nucl. Phys. \textbf{B237} (1984) 397. 
\bibitem{Nariai} H.~Nariai,
``On a new cosmological solution of Einstein's field equations of gravitation'',
Sci. Rep. T\^ohoku Univ. \textbf{35} (1951) 46. 
Gen. Rel. Grav. \textbf{31} (1999) 963. 
\bibitem{WA} S.~Wada and T.~Azuma,
``De Sitter metric as a self-consistent solution of the back reaction problem'', 
Phys. Lett. \textbf{B132} (1983) 313. 
\bibitem{AW} T.~Azuma and S.~Wada,
``Classification of spatially flat cosmological solutions in the presence of
the cosmological constant and backreaction of conformally invariant quantum
fields'',  
Prog. Theor. Phys. \textbf{75} (1986) 845.


\end{thebibliography}


\end{document}